# Modelling and Simulation of Charging and Discharging Processes in Nanocrystal Flash Memories During Program and Erase Operations

## Andrea Campera [*] and Giuseppe Iannaccone

*Dipartimento di Ingegneria dell'Informazione: Elettronica, Informatica, Telecomunicazioni*

*Università di Pisa, Via Caruso 16, 56122, Pisa, Italy*

[*] corresponding author. Tel : +39 0502217639, fax: +39 0502217522, E-mail address: andrea.campera@iet.unipi.it

**Abstract**

This work is focused on the understanding of charging and discharging processes in silicon nanocrystal flash memories during program and erase operations through time-dependent numerical simulations. Time dependent simulations of the program and erase operations are based on a description of the nanocrystal memory dynamics in terms of a master equation. The related transition rates are computed with a one dimensional Poisson-Schrödinger solver which allows the computation of the tunnelling currents and of generation and recombination rates between the outer reservoir and localized states in the dielectric layer. Comparison between simulations and experiments available in the literature provides useful insights of the storing mechanisms. In particular, simulations allow us to rule out that electrons are stored in confined states in the conduction band of silicon nanocrystals, whereas they suggest that electrons are actually trapped in localized states in the silicon gap at an energy close to the silicon valence band edge, and located at the interface between the nanocrystals and the surrounding silicon oxide.

## 1. Introduction

Nanocrystal Flash Memories (NFMs) have been proposed as a promising alternative to conventional Flash Memories [1-3]. The discrete nature of the storage elements significantly reduces the impact of Stress-Induced Leakage Currents (SILCs) on data retention [3-4], overcoming the scaling limits of conventional Non-Volatile-Memories (NVMs). Moreover it makes possible the use of a thinner tunnel oxide, which in turn implies higher programming current, lower program/erase times, and at the same time lower program and erase voltages, with respect to conventional Flash Memories. One of their main drawbacks, up to now, is the relatively small threshold voltage shift achieved and the dispersion of electrical properties due to the spread in dot size and density [3].

Program operations in NFMs can be performed, in general, via hot carrier injection or by means of tunnelling from the channel (Fowler–Nordheim or direct tunnelling). Channel tunnelling mechanisms are preferable for uniform charging and discharging of nanocrystals and for low-power applications, since the tunnelling current is very low. Indeed, as well known, hot carrier injection, and therefore trapped charge, is highly localized at the drain side of the channel. In the rest of this work we will focus on channel tunnelling mechanisms, although channel hot injection provides lower programming times and voltages and would be indispensable for dual bit operation based on asymmetric charging [5].

A proper understanding of the program and erase operations of nanocrystal memories is crucial in order to improve the performance of such memory architectures. Up to this moment various models have been proposed, each of them with the purpose of explaining the Program and Erase curves of NFMs. We can distinguish two different approaches: a floating gate-like approach and a trap-like approach. The former uses a modified version of the model used for Floating Gate Flash Memories, practically computing the dot charge dynamic as a function of the tunnel current density injected in and emitted by the storage layer (see for example [4,6]). The basic equation of such a model is the following:

$$\frac{dQ}{dt} = -J_{e,in} + J_{e,out} + J_{h,in} - J_{h,out} \qquad (1)$$

where Q is the charge density in the nanocrystal, $J_{e,in}$ and $J_{e,out}$ are the tunnel current densities of injected and emitted electrons, whereas $J_{h,in}$ and $J_{h,out}$ are the tunnel current densities of injected and emitted holes. The tunnelling currents can be computed either with the WKB approximation or in a more rigorous way. Such a model has the advantage of the extreme simplicity but does not address the specific nanoscale properties of the dot, namely the presence of discrete states due to strong quantum confinement, the effects of single electron charging, and the appropriate distribution function of electrons in the dot (the small number of electrons challenges the use of the Fermi-Dirac distribution function).

The trap like approach consists in the computation of the generation and recombination rates which are the transition rates from a Kohn-Sham state in the dot to an electronic state in one band of one reservoir and vice versa, respectively. Once the generation and recombination rates have been computed, we can calculate the average number of electrons in the dot, by means of a master equation. This *modus operandi* can provide additional information with respect to the floating gate approach. Indeed it allows us to consider that electrons could not necessarily be stored in confined states in the conduction band [7,8], whereas they could be stored in localized states at the nanocrystal/dielectric interface. It has been proposed that to explain the observed long retention times of NFMs electron storage in deep traps inside the silicon band gap must be considered [9]. The spatial and energetic position and nature of these traps has not yet been identified with certainty.

The issue is obviously relevant also from a technological point of view, since it determines which aspects must be more closely addressed for improving data retention and for reducing program/erase times. In this paper, we investigate such issue by means of time-dependent numerical simulations, based on the solution of a master equation, and demonstrate that only the mechanism according to which electrons are stored in traps localized in the silicon gap at the nanocrystal/surrounding dielectric interface is compatible with the experimental results, such as for example those presented in Ref. [3].

It is always very hard to draw drastic conclusions on an experiment by means of a numerical simulation, which is based on a model that must obviously be rather simplified and idealized with respect to the actual system. Nevertheless, we shall show that results are rather robust with respect to varying parameters as, for instance, the thicknesses of the various layers, and to physical parameters, and enable us to claim with some confidence that electrons are not stored in the nanocrystal conduction band.

The rest of this paper is organized as follows: in Section 2 we will present the main physical model, in Sections 3 and 4 the results on the program and erase operations, respectively. In section 5 we will present a Discussion and then we will draw our Conclusion.

**2. Physical model**

The nanocrystal memory has been approximated by a one-dimensional structure, corresponding to the vertical cross section along the central axis of one nanocrystal (z direction), as shown in figure 1.

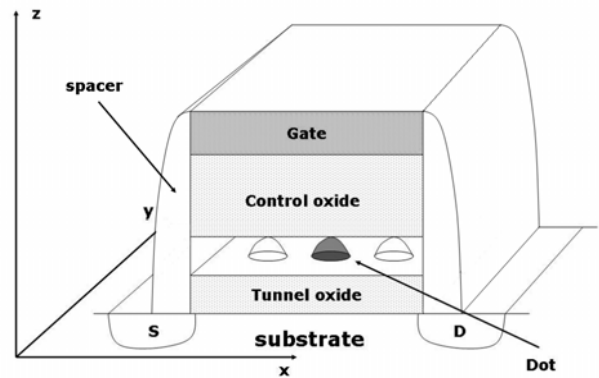

Fig. 1: schematic view of the nanocrystal Flash memory (not in scale). Dots are semispherical because this is the shape most close to the actual one, if dots are deposited by LPCVD.

For the investigation of nanocrystal charging and discharging process we have modified a one-dimensional model [10], that was originally proposed for the study of SILCs in MOS capacitors, which treats tunnelling into and from the nanocrystals in terms of generation and recombination processes. As we have mentioned in the introduction, a complete model describing the dot dynamics should include the possibility to account for two very different cases from the physical point of view:

a) In the first case, we have a quantum dot defined by the electron confinement due to the large gap of surrounding silicon oxide layer. The electrochemical potential of the dot is a function of the bias conditions and of the number of electrons in the dot.
b) The second case corresponds to various localized traps placed inside or at the silicon/silicon oxide interface of each nanocrystal. Let us assume that Coulomb repulsion allows at most one electron per trap. In this case, we can still describe the nanocrystal as a unique system, its electrochemical potential being the minimum energy required to place an electron in one of the traps. For simplicity, we assume that all energy traps are identical, and that when an electron is added to the system, the energy of all trap states increases by a given charging energy.

Here we briefly report the basic aspects of the model, further details on the computation of the tunneling rates are available in [10]. Let us stress our simplifying assumptions:
i) the nanocrystals are identical and uniformly distributed,

*ii)* charged dots do not interact with each other or, in other words, no effect related to capacitive coupling among nanocrystals is taken into account,

*iii)* when an electron is added to the nanocrystal system, its electrochemical potential increases by a factor $\Delta\mu = q^2/2C$, where $q$ is the electron charge and $C$ is the total nanocrystal capacitance.

The total capacitance $C$ can be approximated by:

$$C = C_1 + C_2 = \frac{\pi\varepsilon_{Si}d^2}{2t_1 + \frac{\varepsilon_{ox}}{\varepsilon_{Si}}d} + \frac{\pi\varepsilon_{Si}d^2}{2t_2 + \frac{\varepsilon_{ox}}{\varepsilon_{Si}}d} \quad (2)$$

where $d$ is the nanocrystal diameter, $t_1$ is the tunnel oxide thickness, and $t_2$ is the control oxide thickness, and $\varepsilon_{Si}$ ($\varepsilon_{ox}$) is the silicon (silicon oxide) dielectric constant. We want to remark that, in particular, if electrons are localized in trap states, which are strongly localized states, their ionization energy does not be changed by the Coulombic charging energy, whereas the total trap energy $E_t$ is raised by the Coulombic energy.

We can refer to the 1D profile of conduction and valence bands along the central axis of one nanocrystal (z direction), as shown in Fig. 2. For each applied gate voltage, the band profile is computed with a self-consistent Poisson-Schrödinger solver, which takes into account quantum confinement at the emitter, mass anisotropy in silicon conduction band and light and heavy holes. The computation is performed with the quasi equilibrium approximation, i.e. assuming that the tunneling current is so low that the oxide separates two regions in local equilibrium with two different Fermi energies. Let us also consider the electrochemical potential of the nanocrystal $\mu_{NC}$.

We call generation rate the transition rate from an electrode to the nanocrystal, and recombination rate the transition rate from the nanocrystal to one electrode. As we can see in Fig. 2 there are eight different generation and recombination rates, with an obvious meaning of the terms. Let us now consider a state $|\beta\rangle$ in one band of one electrode, in contact to a (Kohn-Sham) state $|\alpha\rangle$ in the dot.

We can write the transition rate from $|\beta\rangle$ to $|\alpha\rangle$ according to the Fermi "golden rule", as:

$$\nu_{\beta\to\alpha} = \frac{2\pi}{\hbar}|M(\beta,\alpha)|^2 h_\Gamma(E_\alpha - E_\beta) \quad (3)$$

where we have take into account for inelastic transition by replacing the conventional Dirac function with a Lorentzian function, which is expressed by:

$$h_\Gamma(E_\alpha - E_\beta) = \frac{\Gamma/\pi}{(E_\alpha - E_\beta)^2 + \Gamma^2} \quad (4)$$

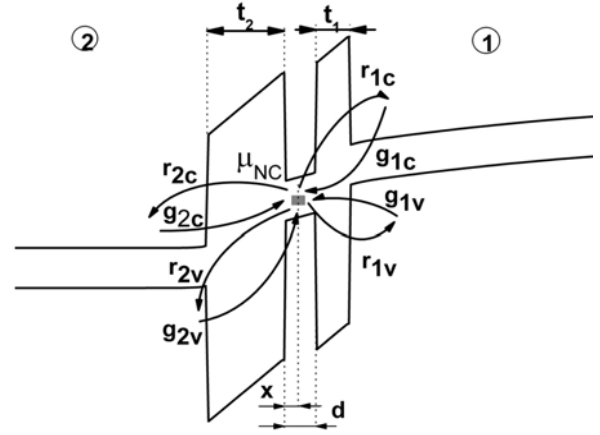

Fig. 2: Band profile of the structure used in this work and generation and recombination rates. $\mu_{NC}$ indicates the electrochemical potential of the nanocrystal.

The larger $\Gamma$, half width of the Lorentzian function, the larger degree of inelastic transitions are permitted. The transition rate from $|\beta\rangle$ to $|\alpha\rangle$ can also be rewritten as:

$$\nu_{\beta\to\alpha} = \sigma_{\beta,\alpha}J(\beta,x') = \sigma_{\beta,\alpha}T(E_l)\nu(E_l) \quad (5)$$

Comparing (3) with (5) we can define a trap cross section per unit energy as:

$$\sigma_{\alpha,\beta} = k \cdot h_\Gamma(E_\alpha - E_\beta) \quad (6)$$

where $k$ is an unknown constant surface (whose dimensions are m$^2$), to be determined via fitting with experiments. This definition of capture cross section is slightly different from the conventional one, in the fact that a dependence on the energy difference between initial and final states is introduced. The computation of all generation and recombination rates can therefore be done with a unique unknown multiplying coefficient, i.e. $k$. By setting for the moment $k$ to 1 m$^2$, all time quantities will be obtained in units of seconds divided by $k$.

The general expressions for the generation and recombination rates are obtained integrating the transition rate from $|\beta\rangle$ to $|\alpha\rangle$ all over the possible occupied or unoccupied states in the reservoir, respectively:

$$g = 2\int_\beta \nu_{\beta\to\alpha}\rho_\beta f_\beta d\beta \quad (7)$$

$$r = \int_\beta \nu_{\beta\to\alpha}\rho_\beta(1-f_\beta)d\beta \quad (8)$$

The factor 2 in the generation rate takes into account for the two possible spin states for such a mechanism, whereas during the recombination process the final state must have the same spin of the trapped electron. As an example we report the extended expression for the generation rate from the conduction band of the substrate to the dot:

$$g_{1c} = 2\int_{\beta} \sigma_{\beta,\alpha} T(E_l) \nu(E_l) \rho_{\beta}(E) f_{\beta}(E) d\beta =$$

$$= 2\int_{E_l}^{\infty} k \cdot T(E_l) \nu(E_l) \rho_1(E_l) \rho_T dE_l \quad (9)$$

$$\times \int_0^{\infty} h_T(E_l + E_T - E_\alpha) f_1(E_l + E_T) dE_T$$

Let us indicate with $g(n)$ the total generation rate of the n-th electron, i.e. the probability per unit time that the n-th electron is added to the nanocrystal summed over all possible transitions. Similarly, we define $r(n)$ as the total recombination rate of the n-th electron. If $P(n,t)$ is the probability that $n$ electrons are in the nanocrystal at time $t$, we can write the following master equation:

$$\frac{dP(n,t)}{dt} = r(n+1)P(n+1,t) + g(n)P(n-1,t) - P(n,t)[r(n) + g(n+1)] \quad (10)$$

During the charging process (program operation), equation (10) is numerically solved with the following boundary conditions:

$$\begin{cases} P(n,0) = 0 & \forall n \geq 1 \\ P(0,0) = 1 \end{cases} \quad (11)$$

corresponding to zero electron in the dot at time zero. We can therefore obtain the average number of electrons in the dot as a function of time as

$$\langle n(t) \rangle = \sum_{n=0}^{\infty} nP(n,t) \quad (12)$$

and the average threshold voltage shift as a function of time as

$$\langle \Delta V_T(t) \rangle = \gamma \langle n(t) \rangle \quad (13)$$

where

$$\gamma = \frac{q\rho_{NC}}{\varepsilon_{ox}} \left( \frac{d\varepsilon_{ox}}{4\varepsilon_{Si}} + t_2 \right) \quad (14)$$

For the simulation of the discharging process (erase operation) the boundary conditions are:

$$\begin{cases} P(m,0) = 1 \\ P(n,0) = 0 & \forall n < m \end{cases} \quad (15)$$

where $m$ is the initial number of electrons inside the dot, that we choose so that $m\gamma$ is equal to the initially programmed threshold voltage.

## 3. Program operation

We apply our model to typical structures for which experiments of program/erase operations are available. In particular, we show the results that we have obtained considering the structure indicated with B in Ref. [3]: it has control oxide thickness of 8 nm, tunnel oxide thickness 3.5 nm, average nanocrystal diameter 3.2 nm, nanocrystal density $2 \times 10^{11}$ cm$^{-2}$. The experimental threshold voltage shift is plotted as a function of the write voltage for different times, and for program and erase operations, in Fig. 3, where we report data of Fig.7 of Ref. [3], device B.

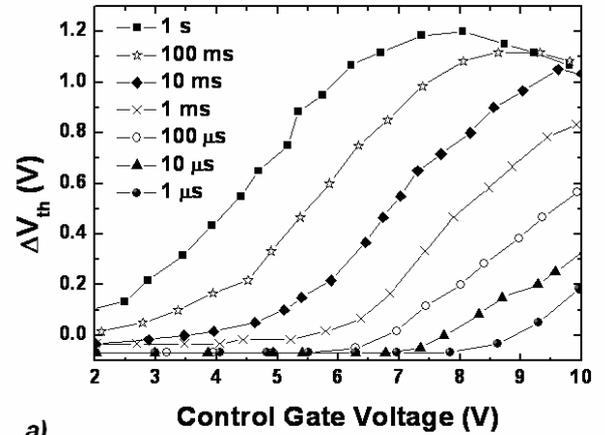

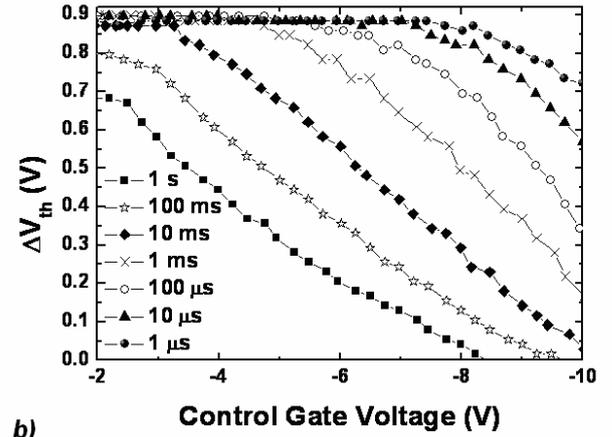

Fig. 3: Here we reproduce experimental results extracted from Fig. 7 of Ref. [3], for the device indicated with B. In Fig. 3 *a)* the experimental programming curves are shown, while in Fig. 3 *b)* the erasing curves are reported.

In the simulations, we consider a tunnelling effective mass of electrons in silicon oxide of 0.5 $m_0$, and of holes in such a material of 0.4 $m_0$. Let us first consider the results from a simulation performed considering the electrons in the conduction band of the nanocrystal. As we can see in Fig. 4, simulations are very different from experiments: first of all, the peak of $\Delta V_T$ is obtained for gate voltages smaller than 4 V for all program times, while we can see in Fig. 3 that in experiments the peak of $\Delta V_T$ is obtained for gate voltages larger than 7 V. As a second point we can observe that the simulated programming window is very small with respect to the experimental one, this fact is caused by the high recombination rate $r_{2c}$ towards the gate electrode since electrons transmission coefficient is small through the triangular barrier.

This behavior is very robust with respect to variations of the values of meaningful parameters of the model, i.e. if the thickness of the various layers or the physical parameters of the materials are changed, simulation results are similar and always very far from experiments.

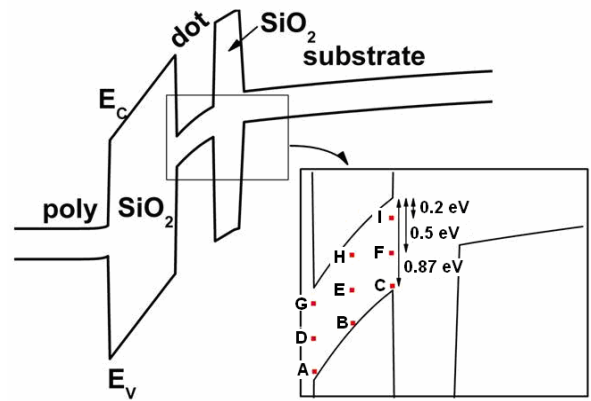

Fig. 5: Band profiles and (inset) nine trap positions and energies considered in the simulation.

The simulated program characteristics for traps located in the center of the dot layer (trap B of Fig. 5) and with an energy level of 0.87 eV below the dot conduction band are shown in Fig. 6.

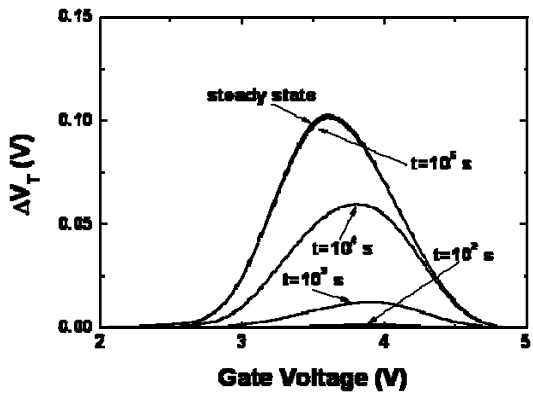

Fig. 4: Average threshold voltage shift obtained considering electrons in the conduction band of the dot. The reader should pay attention in particular to the shape and the programming window, not yet to the programming time.

Let us stress the fact that the programming times reported in Fig. 4 are obtained by considering the best fitting value for the constant *k* which have been extracted from the comparison between simulations and experiments. However, Fig. 4 shows that for any *k* the conclusions would be similar.

Given such huge disagreement with experiments, we have investigated the program/erase behaviour for a series of trap positions and single electron energy levels, in particular shown in the inset of Fig. 5. Here we do not have the possibility to show all results: however, results in qualitative accord with experiments are obtained for trap energies from about 0.8 eV to 1 eV below the nanocrystal conduction band and positioned in the centre of the nanocrystal layer or towards the control oxide.

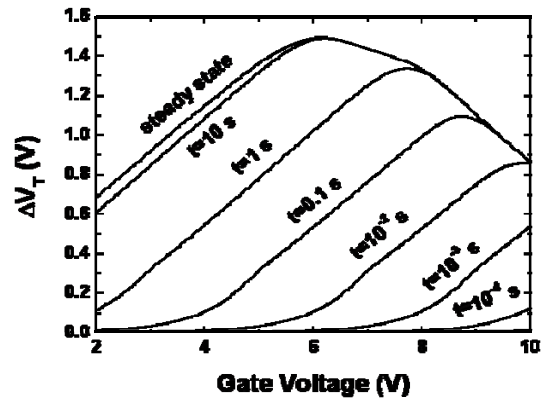

Fig. 6: Average threshold voltage shift versus gate programming voltage for different write times. The trap energy is 0.87 eV below the CB and is placed in the center of the dot (trap B of Fig. 5).

A qualitative agreement between theory and experiments (Fig. 3*a*) can be noticed, therefore we have extracted from this simulations the value for the constant *k*, which is resulted to be 300. We want to remark again that we are not looking for a quantitative agreement with experiments since a 1-D model can not be extremely accurate. In Fig. 7 the simulation results that we have found placing traps 0.87 eV below the dot conduction band and at the nanocrystal/control oxide interface are reported. We can still observe a good agreement with experiments as far as programming window and programming times are concerned.

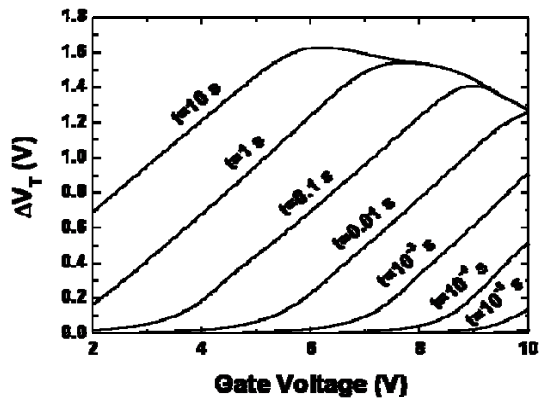

Fig. 7: Average threshold voltage shift versus gate program voltage for different write times. The trap correspond to position and energy indicated with A in Fig. 5.

Characteristics exhibit very evident qualitative changes as we modify trap position and energy, which cannot be recovered by simply adjusting the model parameters. In Fig. 8 we can see the results we obtain if we consider trap C of Fig. 5 still 0.87 eV below the silicon conduction band edge.

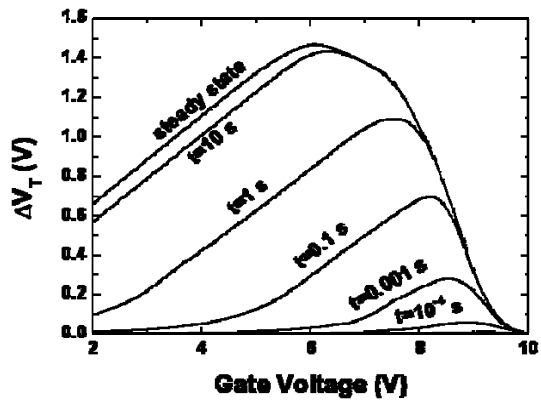

Fig. 8: Average threshold voltage shift versus gate programming voltage for different write times. The trap correspond to energy and position indicated with C in Fig. 5.

The shape of the programming curves is now very different from the previous two and hence from experiments. If we change the trap energy what we have observed is that the programming times begin to increase, whereas the programming windows begin to decrease. As a general example we report in Fig. 9 the results obtained considering electrons in the center of the dot and 0.5 eV below the silicon conduction band edge (type D of Fig. 5).

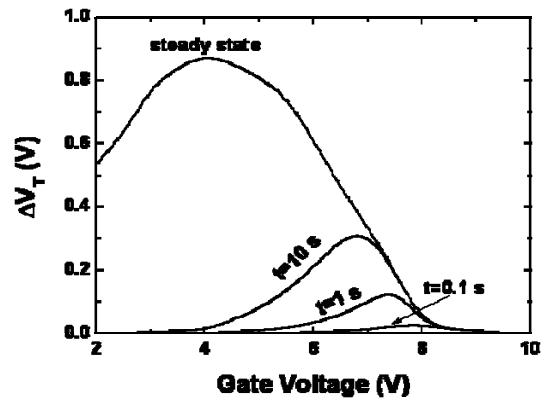

Fig. 9: Average threshold voltage shift versus gate programming voltage for different write times. The trap energy is 0.5 eV below the silicon conduction band edge and in the center of the dot (E of Fig. 5). A qualitative discrepancy, as far as the programming times and the programming window concern, is observed.

Up to now we have not explicitly take into account the actual shape of the dot. While the dot shape is not precisely known, the situation is typically that sketched in Fig. 10, which translates in a different cross section of the nanocrystal states for transitions through the two barriers. Let us introduce a "form factor" $\beta$ defined as the ratio of the cross section for transitions through the control dielectric to the cross section for transitions through the tunnel oxide.

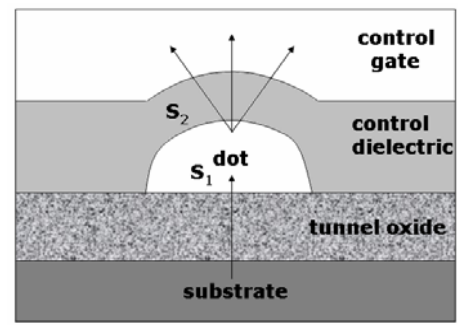

Fig. 10: schematic of the nanocrystal cell used in this work. In this figure is reported a more accurate structure of the cross section and the area of the surfaces involved in the generation and recombination processes.

Nevertheless, we show in Fig. 11 that even if we consider very different values of $\beta$, such as 5 and 1, the simulated program characteristics do not charge from a qualitative point of view. This is due to the fact that the dominant terms are represented by the tunnelling probabilities, which are exponentially dependent on the energy and the applied gate voltage, and only linearly dependent on surfaces.

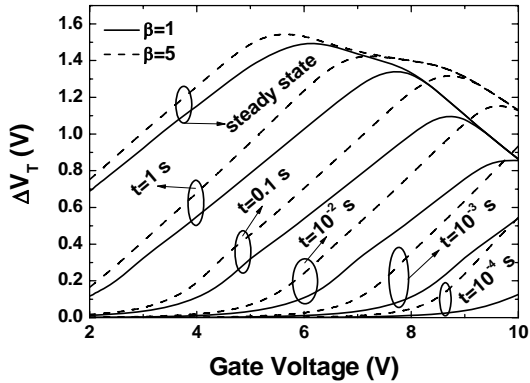

Fig. 11: in this figure are reported the programming curves with and without the shape factor $\beta$. Simulations regards trap center in the center of the dot and 0.87 eV below its conduction band (trap B in Fig. 5).

It is interesting at this point to evaluate the conventional (energy independent) cross section of the traps we are considering responsible for nanocrystal storage. The total average current from a reservoir to a nanocrystal can be rewritten as

$$I_{TOT} = \int \sigma(E) J(E) dE = \overline{\sigma} \int J(E) dE \quad (16)$$

where $\sigma(E)$ is our energy dependent cross section, $\overline{\sigma}$ the "conventional" capture cross section, and $J(E)$ is the current density per unit energy injected through the associated barrier: $\overline{\sigma}$ can hence be obtained as:

$$\overline{\sigma} = \frac{\int \sigma(E) J(E) dE}{\int J(E) dE}. \quad (17)$$

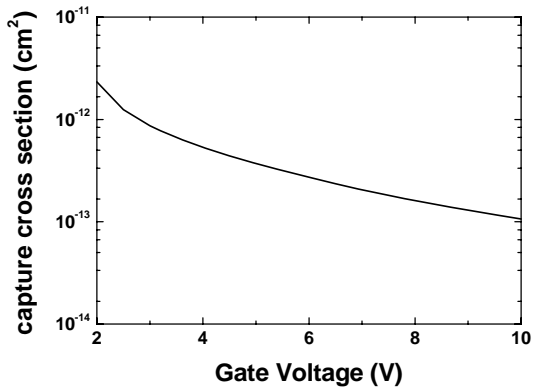

Fig. 12: conventional capture cross section as a function of the gate applied voltage. This figure is extracted for traps in the center of the dot, 0.87 eV below the CB (trap B in Fig. 5) and considering no electron inside the dot.

Of course, if $\sigma(E)$ is a property of the trap and is independent of the applied bias, according to (17) $\overline{\sigma}$ will depend on the applied bias through $J(E)$. In figure 12 we plot the "conventional" capture cross section for a trap placed in the center of the dot (trap B of Fig. 5) and 0.87 eV below silicon conduction band edge: as can be seen, its values are in the range $10^{-12} \sim 10^{-13}$ cm$^2$ (compare with experimental results obtained in [13]).

**4. Erase operation**

As far as the erase operation is concerned, we first observe that if we assume that only direct tunnelling from the traps is involved, we find a result in disagreement with experiments. In order to explain this behaviour we can look at the band profile sketched in figure 13.

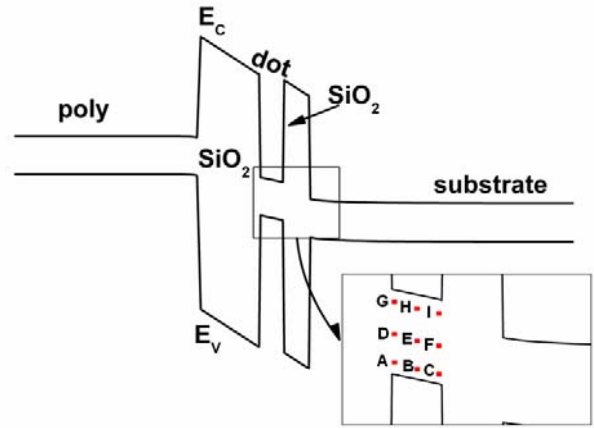

Fig. 13: Band profile of the nanocrystal memory structure during erase operations. In the inset the trap states that we have considered are reported.

In figure 14, the results of simulations obtained considering electron storage in traps located in the center of the dot and 0.87 eV below the nanocrystal conduction band are reported. We can observe that the erase process is too slow compared with experiments in Fig. 3 *b)* and the shape of the erase characteristics is different.

We then assume that a second discharge mechanism is present, a two-step process involving the thermally de-trapping of electrons to the conduction band and then the recombination towards the substrate. If we assume that the thermal emission of electrons from traps to the conduction band is infinitely faster than recombination, we obtain the discharge curves shown in figure 15. Yet again discharge is too fast compared with experiments. If we consider the experimental results for each erase time, from 1 μs to 1 s, we can observe that they are comprised between the characteristics given by discharge from the conduction band and those given by discharge from inner traps, as can be seen for example in Fig. 16 for the erase time of 1 ms. It is then possible to assume that both mechanisms concur in the discharge process.

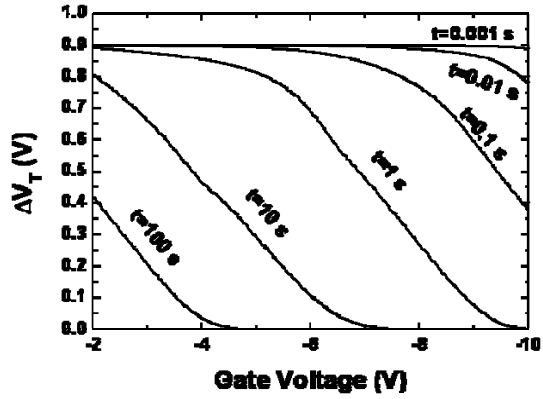

Fig. 14: Threshold voltage shift as a function of the gate voltage for different erase times. Electrons are stored in traps at the center of the nanocrystal layer, 0.87 eV below the silicon conduction band (trap B in fig. 13).

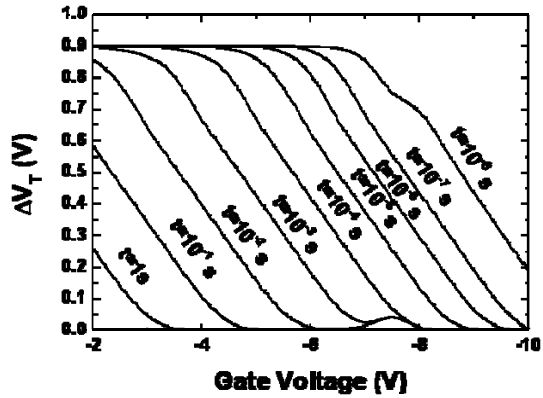

Fig. 15: Threshold voltage shift as a function of $-V_G$ for different erase times. Electrons are located in the nanocrystal conduction band.

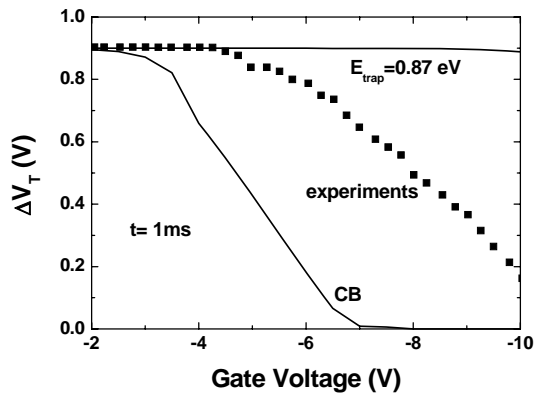

Fig. 16: comparison between experimental data, discharge from the conduction band and discharge from the trap, for an erasing time of 1 ms. Experiment from [3] are included between the two simulated curves also for all the other times.

The relative importance of the two processes depends on the efficiency of the de-trapping mechanism. We can then write a master equation that takes into account two different electron populations in the nanocrystal (electrons in deep traps and in the conduction band) and emission/capture rates between the conduction band and the traps:

$$\frac{dP(N_T, N_{CB}, t)}{dt} = P(N_T+1, N_{CB}, t) r_T(N_T+1)$$
$$+ P(N_T, N_{CB}+1, t) r_{CB}(N_{CB}+1) + P(N_T-1, N_{CB}, t)$$
$$\times g_T(N_T) + P(N_T, N_{CB}-1, t) g_{CB}(N_{CB})$$
$$+ P(N_T+1, N_{CB}-1, t) e(N_T+1, N_{CB}-1) \quad (18)$$
$$+ P(N_T-1, N_{CB}+1, t) d(N_T-1, N_{CB}+1)$$
$$- P(N_T, N_{CB}, t)[r_T(N_T) + r_{CB}(N_{CB}) + g_T(N_T+1)$$
$$+ g_{CB}(N_{CB}+1) + d(N_T, N_{CB}) + e(N_T, N_{CB})]$$

where $e$ is the emission rate between the trap and the ground state in the conduction band, $d$ is the capture rate from the conduction band to the trap and $P(N_T, N_{CB}, t)$ is the probability per unit time that $N_T$ electrons stored in localized traps associated to the nanocrystal and $N_{CB}$ electrons are stored in confined states in the nanocrystal conduction band. Moreover we have indicated with $r_T$, $g_T$ ($r_{CB}$, $g_{CB}$) the total recombination and generation rates of the trap state (conduction band state), respectively.

The following initial conditions must be applied:

$$\begin{cases} P(N_T, N_{CB}, 0) = 1, & \text{if } N_T = N, N_{CB} = 0 \\ P(N_T, N_{CB}, 0) = 0, & \text{otherwise} \end{cases} \quad (19)$$

which means that at time 0 all the electrons are stored in the deep trap states.

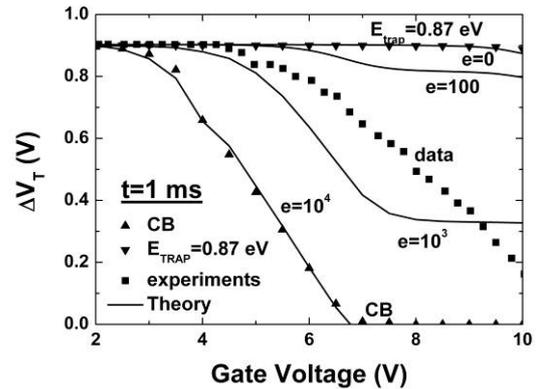

Fig. 17: comparison between experimental data, discharge from the conduction band, discharge from the trap and two level system discharge for different emission rates, for an erasing time of 1 ms.

In figure 17 the results of simulations with different emission rates $e$ are shown. We can observe that a constant emission rate can not explain exactly experiments, but we believe that it allows to better reproduce the experiments. Further work is needed to understand whether the emission rate *e* may depend on the applied voltage for a sort of Stark effect, or whether some additional indirect recombination mechanisms may be relevant.

## 5. Discussion

Now we want to discuss a little bit more in detail the results of our simulations, together with some experiments well known in literature that can provide useful insights. More in particular we can think that if electrons are actually stored in interfacial traps and not in conduction band of the nanocrystals, reducing the number of interfacial traps we should observe a reduction, at least, of the programming window since electrons should now be stored in the conduction band.

Moreover a reduction of the retention time should also be observed. This is what had actually been observed in a work of Shi and coworkers [12]. They noticed that the maximum shift in the C-V hysteresis loop was obtained in a vacuum annealed device (having a high trap density), whereas the minimum shift was obtained in the $H_2$ annealed device (having a low trap density), and the middle in the as-deposited device.
They explained this fact by assuming that more charge was stored in the vacuum-annealed nanocrystals than in those $H_2$-annealed. Moreover they observed that the long retention time was not compatible with the hypothesis that injected electrons were stored dominantly in the conduction band, especially in the case of more than one electron stored in the nanocrystal. This is not an out of the ordinary fact since trapping centers play a critical role also in other memory structures, as for instance SONOS memories where charge are stored in deep traps and mainly at or close to the nitride/oxide interfaces. During program operations the injected electrons will first fill empty states with a deeper trap energy, and then will progressively fill states where the trap energy is shallower. Therefore it is comprehensible the spatial position of traps that we have indicated as responsible for the program operations, indeed they will fill first the traps located at the control dielectric/nanocrystal interface where the trap energies are lower.
The choice of the trap energy, in particular 0.87 eV below the dot conduction band, is related to some experimental results found by Kwon and co-workers [13]. They found that memory effect was dominantly related to hydrogen-related traps, in addition to the three-dimensional quantum confinement and Coulomb charge effects. Deep level transient spectroscopy exposed that the activation energies of the hydrogen-related traps are $E_v+0.29$ eV (H1) and $E_v+0.42$ eV (H2).

## 6. Conclusion

Comparison of numerical simulations of program/erase characteristics of nanocrystal memories with experiments published in the literature allows us to exclude that electrons are stored in quantum confined states in the nanocrystal conduction band. Even if the model used is rather simplified and idealized, the results are robust enough to rule out such possibility.

Let us stress the fact that, notwithstanding its simplicity, the proposed model provides a qualitative and reasonable quantitative agreement with experiments. Further investigations are needed, in order to achieve quantitative agreement with experiments also for the erase operation and for a broader set of experimental structures. The discharge process is very likely due to a concurrence of different recombination mechanisms and need to be further investigated. As a final remark we can say that from the technological point of view we believe that the quality of the interface between the silicon dots and the surrounding oxide is more relevant than the size and the shape of the dots in determining the program and erase behavior.

**Acknowledgements**

The present work has been supported by IEIIT-CNR through the FIRB-MIUR project "Sistemi miniaturizzati per elettronica e fotonica".